\documentclass[twocolumn,showpacs,amsmath,amssymb]{revtex4}
\usepackage{graphicx}% Include figure files
\usepackage{dcolumn}% Align table columns on decimal point
\usepackage{bm}% bold math

\begin{document}

\title{Intrinsic Decoherence and Irreversibility \\ in the Quasiperiodic Kicked Rotor}
\author{A.C. Sicardi Shifino}
\author{G. Abal}
\author{R. Siri}
\author{A. Romanelli}
\affiliation{Instituto de F\'{\i}sica,
Universidad de la Rep\'ublica\\
CC 30, CP 11000, Montevideo, Uruguay}

\author{R. Donangelo}
\affiliation{Instituto de F\'{\i}sica,
Universidade Federal do Rio de Janeiro,\\
C.P. 68528, 21941-972 Rio de Janeiro, Brazil}

\date{\today}

\begin{abstract}
We show that some classically chaotic quantum systems uncoupled from noisy
environments may generate intrinsic decoherence with all its associated effects.
In particular, we have observed time irreversibility and high sensitivity to small
perturbations in the initial conditions in a quasiperiodic version of the kicked 
rotor. The existence of simple quantum systems with intrinsic decoherence
clarifies the quantum--classical correspondence in chaotic systems.
\end{abstract}


\maketitle

It is well established, mainly from the important contributions by Zurek and
others \cite{Zurek}, that weak interaction of a quantum system with a noisy
environment induces decoherence in the quantum dynamical evolution. Ultimately
this effect is behind the emergence of a classical world from the quantum
formalism. In the case of quantum systems with chaotic classical analogs,
this implies that diffusion, entropy increase, time irreversibility and other
typical characteristics of classical chaos emerge as a result from the decoherence
induced by noisy environments. Experimental evidence of the effects of the decoherence
induced by coupling to noisy environments has been obtained in the case of
the quantum kicked rotor using optical atom traps. The effect of decoherence is 
to destroy dynamical localization and establish a quantum diffusion regime \cite{edeco_qkr}.
Similar effects have also been observed using trapped ions \cite{edeco_trapions}.
It is interesting to inquire how essential is the role of
the environment in the emergence of classical-like chaotic behavior in quantum systems.
Can the same dynamical effects due to a noisy environment be obtained from a
modified hamiltonian without coupling to the environment? In other words, can a quantum
system develop ``intrinsic decoherence''?

In this paper we show that a modified version of the kicked rotor
can develop classical--like effects usually associated to
environment--induced decoherence. These effects include irreversibility and
a high sensibility to small perturbations in the initial conditions.
The possibility of the generation of ``intrinsic decoherence'' in systems
isolated from their environment, has recently been suggested in the literature \cite{Castagnino,  Gong}.

We recall that in the classical kicked rotor,
for strong enough kick strengths, there is global chaos in phase space \cite{caos}
with all its manifestations. This includes exponential sensibility to small
perturbations in the initial conditions and time--irreversibility. Despite
that the equations of motion are time--reversible, if time is reversed,
small errors are exponentially amplified and a new diffusive process takes
place. In the quantum mechanical case, the diffusive process in action space is
suppressed after a characteristic time due to dynamical localization,
a quantum coherence effect. The quantum suppression of chaos due to dynamical
localization takes place ultimately due to the discrete nature of the
quasienergy spectrum of the kicked rotor \cite{Izrailev}. If time is reversed,
the unitary evolution returns the system to its initial state, in spite of small
errors in the initial conditions. This striking manifestation of the
quantum suppression of chaos was first shown numerically by Graham and
Dittrich \cite{Graham}.

We now introduce a specific model system, the quasiperiodic kicked rotor.
Consider a rotor to which a periodic series of impulsive kicks 
of strength $K$ and period $T_1$ is applied. If a second periodic series of kicks with the
same strength and a period $T_2$ incommensurate with the first, is applied to the
rotor the resulting time--dependent hamiltonian is quasiperiodic and can be
written as
\begin{equation}
H=\frac{P^{2}}{2I}+K\cos\theta \left[\sum _{n=1}^{\infty }
\delta (t-nT_{1})+\sum _{m=1}^{\infty } \delta (t-mT_{2})\right],
\label{eq:qkr2f_ham}
\end{equation}
where $P$ is the angular momentum, $\theta$ is the angular position of the rotor 
and we choose units so that the moment of inertia is $I=1$.
It has been shown \cite{alejo1} that in this system the initial
diffusion persists for arbitrarily long times. In fact, this system can be described
by a master equation and the entropy increases forever in a way typical of a
decoherent evolution \cite{alejo2}. It is worth noting that the decoherence we are
discussing here is not due to the coupling with a noisy environment, but is due to 
the dynamical complexity of the hamiltonian. When the second series of kicks is added
to the kicked rotor, the underlaying frequency response spectrum changes from
a discrete (pure--point) to a dense spectrum \cite{alejo1}.

It is interesting to find whether this intrinsic decoherence also implies a
time--irreversible quantum dynamics, as it is the case with
environment--induced decoherence. If we reverse time in the quasiperiodic
kicked rotor, we find that there is no sensitivity to small perturbations in
the wavefunction.  However, a closer look reveals that perturbing the wavefunction
components is not analogous to perturbing the classical phase space
coordinates. The classical analog of a perturbation of the wavefunction
would be perhaps to perturb the Liouville function of the corresponding
classical system. As is well known, the linear nature of the Liouville equation
prevents any small perturbation in the distribution function from growing
exponentially fast, even under conditions of global chaos.
The actual analog of perturbing the classical phase space
coordinates, would be to perturb the expected value of some suitable quantum
observable. Then, small perturbations in the expectation value of these
observables may, under certain circumstances, lead to large variations
in the wave function.

We have tested the previous considerations using the quasiperiodic kicked rotor
described by~(\ref{eq:qkr2f_ham}). In the position representation,
the wavefunction is
\begin{equation}
\Psi(\theta,t)~=~\frac{1}{\sqrt{2}}\sum_{\ell=-\infty}^{\infty}a_{\ell}(t)e^{i\ell\theta}.
\label{eq:wavefunction}
\end{equation}
The evolution for $a_\ell(t)$ is obtained in a straightforward way \cite{alejo1} from
the quantum map
\begin{equation}
\label{mapa}
a_{\ell}(t_{n+1})=\sum _{j=-\infty }^{\infty }i^{-(j-\ell)}
e^{-i\hbar j^2\Delta t_{n}/2 }\, J_{j-\ell}(K/\hbar)a_{j}(t_{n}),
\end{equation}
where we refer to the instant immediately after the $n^{th}$ kick
as $t_{n}$, to the time interval between two consecutive kicks as $\Delta
t_{n}\equiv t_{n+1}-t_{n}$ and $J_{k}$ is the $k^{th}$ order cylindrical
Bessel function. The time intervals $\Delta t_{n}$ are in general different, since
they depend on the kick sequence. In fact, they form a dense set. Note that
the periodically kicked rotor is also described by the quantum map (\ref{mapa})
with a constant time interval between kicks $\Delta t=T=1$.

If a small perturbation is introduced in the angular position,
$\theta\rightarrow\theta + \varepsilon$, the wavefunction~(\ref{eq:wavefunction}) becomes

\begin{equation}
\tilde\Psi(\theta,t)~=~\frac{1}{\sqrt{2}}\sum_{\ell=-\infty}^{\infty}a_{\ell}(t)
e^{i\ell\varepsilon}e^{i\ell\theta}=\frac{1}{\sqrt{2}}\sum_{\ell=-\infty}^{\infty}\tilde
a_{\ell}(t)e^{i\ell\theta}.
\label{eq:tr_wavefunction}
\end{equation}
where the transformed coefficients $\tilde a_\ell=a_\ell e^{i\ell\varepsilon}$
have acquired a phase. Note that for given $\varepsilon$, the wave function
components with larger $\ell$ undergo larger phase shifts.

\begin{figure}
\includegraphics[scale=0.5, angle=-90]{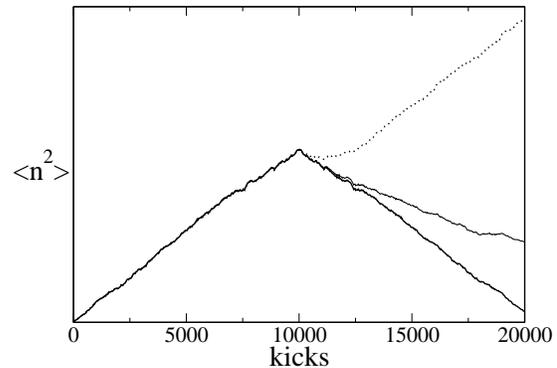}
\caption{\label{fig:eps}
\footnotesize
Effect of different perturbation strengths on the
time evolution of $\left< n^2\right>$ in the quasiperiodic kicked rotor.
Time is inverted after $10^4$ kicks and different perturbation strengths,
$\varepsilon =  10^{-3}$ (thick line),
$\varepsilon =  3\times 10^{-3}$ (thin line) and $\varepsilon =  10\times 10^{-3}$
(dotted line) are applied to the state vector as described in the text.}
\end{figure}

We consider the evolution of an initial wavepacket under the quantum map
(\ref{mapa}).  At a certain point $t^*$ the direction of time is reversed,
$t\rightarrow -t$ and a small perturbation $\varepsilon$ is introduced in
the angular position, as described in~(\ref{eq:tr_wavefunction}).
As shown in figure~\ref{fig:eps}, if the perturbation is above a certain
threshold, the system resumes diffusion after a characteristic time which
depends on $\varepsilon$. This can be understood by considering
eq.~(\ref{eq:tr_wavefunction}) which shows how the perturbation affects the
wavefunction. We adopt the following criteria to decide whether a given angular momentum
component is present in the wavefunction: consider a small number $\delta\ll 1$,
then the angular momentum component $\ell$ is present in the wavefunction if the
condition $|a_\ell|^2>\delta$ is satisfied. We define $\ell_{max}$, as the  highest
angular momentum component present in the wavefunction.
If $\ell_{max}\varepsilon\ll 1$, the perturbation will
have no appreciable effect on the dynamics and the reversibility associated to
a unitary quantum evolution will manifest itself. On the other hand, if
$\ell_{max}\varepsilon\approx 1$, the perturbation affects strongly the wavefunction
and irreversibility will be apparent after a characteristic time. The
weaker the perturbation, the larger this characteristic time, as we show in
figure~\ref{fig:eps}.  For a weak enough perturbation the system will
return to its initial state in a way characteristic of time--reversible systems.
The threshold value of the perturbation is simply related to the maximum angular
momentum present in the wavefunction
\begin{equation}
\varepsilon_{th}\approx\frac{1}{\ell_{max}}.
\label{eq:threshold}
\end{equation}

In this system, as time progresses, diffusion in angular momentum causes progressively
higher angular momentum modes to participate in the dynamics, in the sense
detailed above. Therefore, the sensitivity of the wavefunction to a given
perturbation, as measured by $\varepsilon_{th}$, increases with time.
% FIGURE 1 - epsilon

\begin{figure}
\includegraphics[scale=0.8,angle=-90]{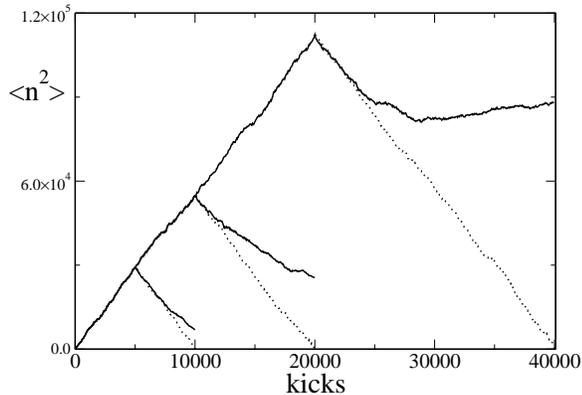}
\caption{\label{fig:tstar}
\footnotesize
Effect of inverting time at different points in the evolution of $\left< n^2\right>$ for
the quasiperiodic kicked rotor. Time is inverted after $5000$ kicks, $10000$ kicks and $20000$ kicks.
The same perturbation $\varepsilon = 3\times 10^{-3}$ is applied in all cases.
The dotted lines represent the unperturbed reversible evolution after time reversal.}
\end{figure}

In figure~\ref{fig:tstar} we show this increasing sensibility by
considering the effect of a given perturbation for different ``break times''
$t^*$. The time the system takes to resume diffusion decreases as $t^*$ becomes
larger, indicating an increasing sensibility to the same perturbation. This is
due to the fact that the threshold value $\varepsilon_{th}$ decreases with time as
$\ell_{max}$ increases. This diffuse process continues for arbitrary long
times, in the quasiperiodic kicked rotor, a system with a dense frequency
response spectrum. That is, the frequencies involved in a Fourier analysis of the
dynamical response of the system form a dense set \cite{alejo1}. (For
non-periodic time--dependent hamiltonians the frequency response spectrum is
its natural generalization of the quasienergy spectrum of periodic systems).

% FIGURE 2 - tstar

It is instructive to consider the case of a discrete spectrum, as that of the
periodically kicked rotor. If we consider the effect of time reversal and a perturbation on
this system, it is not surprising that for reversal times $t^*$ less than the
localization time $\tau$, we find that figures~\ref{fig:eps} and
\ref{fig:tstar} apply also to this system and there is a threshold
$\varepsilon_{th}$ for reversible perturbations. However, in
this case the threshold decreases during the quantum diffusion stage, but stops
decreasing as soon as the wavefunction becomes exponentially localized because
after this point no new angular momentum components are introduced.
Therefore, a perturbation $\varepsilon<\varepsilon_{th}$ which takes place at
any time after the wave function has localized, will not affect
the reversibility of the motion. A side effect of dynamical localization
is to ``freeze" the sensitivity to perturbations at the value it had at time
$t=\tau$ when the system localized.

In conclusion, we have analyzed an example of a classically chaotic quantum
system, the quasiperiodic kicked rotor, which has a decoherent quantum dynamics
without coupling with a noisy environment.
We have shown that suitable small perturbations in the initial conditions are
amplified in a way similar to the classical case and as a result, the quantum
motion becomes irreversible. It should be stressed that by
``suitable perturbations'' in the quantum mechanical context, we mean a
small perturbation in the expected value of an observable and not a small
perturbation in the wave function. As we have shown, a small perturbation of the
mean value of a quantum observable may imply a large perturbation of the
wave function.

In the case of the periodically kicked rotor, the dynamical localization
prevents a similar effect to take place. The quasiperiodic
kicked rotor has a dense response spectrum \cite{alejo1} while the periodically
kicked rotor has a discrete spectrum which produces localization. Thus we
conjecture that a dense spectrum is necessary in order to observe
intrinsic decoherence in classically chaotic quantum systems.
We note that this requirement has also been proposed in a cosmological context
\cite{Castagnino}.

The existence of simple quantum systems with intrinsic decoherence and
the associated time irreversibility and high sensitivity to small perturbations
in initial conditions may help to clarify the quantum--classical correspondence
in chaotic systems.

\emph{We acknowledge the support of PEDECIBA (Uruguay).
GA and RD acknowledge partial financial support from the Brazilian Millennium Institute
for Quantum Information---CNPq (Brazil).}

\end{document}